\documentclass[sigconf,screen,natbib=false,nonacm]{acmart}
\AtBeginDocument{%
  }

\usepackage{fvextra}                 
\usepackage[english]{babel}          
\usepackage[autostyle]{csquotes}     
\MakeOuterQuote{"}                   
\usepackage{xspace}                  
\usepackage[normalem]{ulem}          
\usepackage{soul}                    

\usepackage{mathtools}               
\usepackage{dsfont}                  
\usepackage{bm}                      
\usepackage{siunitx}                 
\usepackage[dvipsnames, svgnames]{xcolor} 
\usepackage{tabularray}              
\UseTblrLibrary{amsmath,siunitx,diagbox,varwidth,functional,counter}             
\SetTblrTemplate{note}{inline}
\SetTblrTemplate{caption}{empty}
\SetTblrStyle{note}{font=\footnotesize}

\usepackage[inline]{enumitem}                

\graphicspath{{./figures/}}          

\usepackage[most]{tcolorbox}         
\tcbuselibrary{all}

\theoremstyle{plain}

\theoremstyle{definition}

\theoremstyle{remark}

\usepackage{balance}                 
\usepackage{footnotehyper}           

\usepackage[capitalize,noabbrev,nameinlink]{cleveref}
\crefname{figure}{Figure}{Figures}
\crefname{appendixfigure}{Appendix figure}{Appendix figures}
\crefname{appendixtable}{Appendix table}{Appendix tables}
\crefname{appendixsubsection}{Appendix \S}{Appendix}
\Crefname{appendixsubsection}{Appendix Subsection}{Appendix Subsections}
\crefname{appendixlisting}{Appendix listing}{Appendix listings}
\Crefname{appendixlisting}{Appendix Listing}{Appendix Listings}

\makeatletter
\@ifclasswith{acmart}{authordraft}{
    \PackageWarning{MyMacros}{ACM Author Draft mode detected!}
    \setlength{\marginparwidth}{2cm} 
    \setlength{\marginparsep}{0cm}   
    \usepackage[obeyDraft,obeyFinal,colorinlistoftodos,textwidth=1.9cm]{todonotes}
    \setuptodonotes{inline}

    \newcommand\redtodo[1]{\textcolor{red}{#1}}
    
    \newcommand\removed[1]{\textcolor{gray}{\sout{#1}}}
    \newcommand\earl[1]{\textcolor{violet}{\footnotesize #1 -- {\em Earl.}}}
    \newcommand\prem[1]{\textcolor{red}{\footnotesize #1 {\em - Prem.}}}
    \newcommand\claudio[1]{\textcolor{ForestGreen}{\footnotesize #1 {\em - Claudio.}}}
    
    \newcommand{\basicalert}[2]{\fbox{\bfseries\sffamily\scriptsize\color{blue} #1}{\sf\small$\blacktriangleright$\textit{\color{red} #2}$\blacktriangleleft$}}

}{
    \@ifpackageloaded{todonotes}{
        \setuptodonotes{disable}
    }{
        \newcommand\todo[1]{}
    }
    \usepackage{environ}

    \newcommand\redtodo[1]{}
    \NewEnviron{redtodos}{} 
    \newcommand\removed[1]{}
    \newcommand\earl[1]{}
    \newcommand\prem[1]{}
    \newcommand\claudio[1]{}
    \newcommand{\basicalert}[2]{}

}{}
\makeatother

\newcounter{rqcounter}
\renewcommand{\therqcounter}{RQ~\arabic{rqcounter}}

\newtcolorbox{rqbox}[1][]{%
  colback=gray!20, colframe=black, fonttitle=\bfseries, fontupper=\small,
  top=0pt, bottom=0pt, left=0pt, right=0pt, #1
}

\newtcolorbox{rqbox1}[2][]{%
  enhanced,
  breakable,
  sharp corners,
  boxrule=0pt,
  frame hidden,
  borderline west={4pt}{0pt}{gray!90}, 
  colback=gray!10,                            
  coltitle=black,                    
  fonttitle=\bfseries,
  attach title to upper,                     
  after title={:\ },                       
  left=3mm, right=3mm, top=1mm, bottom=1mm,
  title={#2},                               
  #1
}
\newcommand{\rqs}[1]{ 
  \refstepcounter{rqcounter}
  \begin{rqbox1}{\therqcounter}
    #1
  \end{rqbox1}
}

\NewDocumentEnvironment{answerenv}{m +b}
{
  \begin{tcolorbox}[enhanced,
  left=3mm, right=3mm,
    colback=gray!10, colframe=gray!80, boxrule=0pt,
    borderline west={4pt}{0pt}{gray!90}]
    \textbf{Answer for RQ#1:} #2
  \end{tcolorbox}
}{}

\newtcolorbox[auto counter, number within=section]{promptbox}[2][]{%
    colback=gray!5,
    colframe=gray!50,
    fonttitle=\bfseries\small,
    title=#2,
    label={#1},
    breakable,
    enhanced,
    coltitle=black,
    boxrule=0.5pt,
    after skip=0pt,
    after=\vspace{-5pt},
    top=1mm,
    bottom=1mm,
    left=1mm,
    right=1mm,
    fontupper=\scriptsize\ttfamily,
}

\newcommand{\placeholder}[1]{%
  \colorbox{blue!10}{\textbf{\texttt{#1}}}%
}

\definecolor{wikiBlue}{HTML}{0645AD}

\NewDocumentCommand{\sref}{m}{\hyperref[#1]{\S~\ref*{#1}}} 

\NewDocumentCommand{\pct}{O{2} O{places} m}{%
  \qty[round-mode=#2, round-precision=#1, round-pad=true]{#3}{\percent}%
}

\NewDocumentCommand{\numrnd}{O{2} O{places} m}{%
  \num[round-mode=#2, round-precision=#1, round-pad=true]{#3}%
}

\NewDocumentCommand{\intnum}{O{} m}{%
  \num[round-mode=none, drop-zero-decimal=true, #1]{#2}%
}

\NewDocumentCommand{\rdel}{}{\ensuremath{\mathcal{R}_\Delta}\xspace} 

\NewDocumentCommand{\pvalues}{}{\textit{p}\nobreakdash-values\xspace}

\NewDocumentCommand{\cruxevalExc}{}{\ensuremath{\textsc{CruxEval}_{\text{exc}}}\xspace}

\NewDocumentCommand{\cruxeval}{}{\textsc{CruxEval}\xspace}

\NewDocumentCommand{\cruxevalMpt}{}{\ensuremath{\textsc{CruxEval}_{p}}\xspace}

\NewDocumentCommand{\cruxevalO}{}{\ensuremath{\textsc{CruxEval}_{o}}\xspace}

\NewDocumentCommand{\cruxevalRenamed}{}{\ensuremath{\textsc{CruxEval}_{v}}\xspace}

\usepackage{listings}

\NewDocumentCommand{\pymint}{ m }{%
  \lstinline[breaklines=true]{#1}%
}

\ExplSyntaxOn

\msg_new:nnnn { ClaudioMacros } { double-period }
  { Double~period~detected~after~#1 ! }
  { You~typed~"#1."~but~the~macro~already~includes~its~own~dot. }

\msg_new:nnnn { ClaudioMacros } { double-comma }
  { Double~comma~detected~after~#1 ! }
  { You~typed~"#1,"~but~the~macro~already~includes~its~own~comma. }

\cs_new_protected:Npn \claudio_check_dot:n #1
  {
    \peek_meaning:NTF .
      { \msg_error:nnn { ClaudioMacros } { double-period } { #1 } }
      { \xspace }
  }

\cs_new_protected:Npn \claudio_check_comma:n #1
  {
    \peek_meaning:NTF ,
      { \msg_error:nnn { ClaudioMacros } { double-comma } { #1 } }
      { \xspace }
  }

\cs_new_protected:Npn \claudio_smart_dot:
  {
    \peek_meaning:NTF .
      { \peek_after:Nw \use_none:n }
      { \xspace }
  }

\cs_new_protected:Npn \claudio_abbr_etc:  { \emph{etc.} \claudio_smart_dot: }
\cs_new_protected:Npn \claudio_abbr_etal: { \emph{et~al.} \claudio_smart_dot: }
\cs_new_protected:Npn \claudio_abbr_vs:   { \emph{vs.} \claudio_smart_dot: }

\cs_new_protected:Npn \claudio_abbr_viz:  { \emph{viz.} \claudio_check_dot:n { \viz } }
\cs_new_protected:Npn \claudio_abbr_wrt:  { w.r.t. \claudio_check_dot:n { \wrt } }

\cs_new_protected:Npn \claudio_abbr_ie:   { \emph{i.e.}, \claudio_check_comma:n { \ie } }
\cs_new_protected:Npn \claudio_abbr_eg:   { \emph{e.g.}, \claudio_check_comma:n { \eg } }

\NewDocumentCommand{\etc}{}{ \claudio_abbr_etc: }
\NewDocumentCommand{\etal}{}{ \claudio_abbr_etal: }
\NewDocumentCommand{\vs}{}{ \claudio_abbr_vs: }
\NewDocumentCommand{\viz}{}{ \claudio_abbr_viz: }
\NewDocumentCommand{\wrt}{}{ \claudio_abbr_wrt: }
\NewDocumentCommand{\ie}{}{ \claudio_abbr_ie: }
\NewDocumentCommand{\eg}{}{ \claudio_abbr_eg: }

\ExplSyntaxOff

\newsavebox{\ghostbox}

\NewDocumentCommand{\qwenMathOneB}{}{\textsc{Qwen2.5}-\textsc{Math}-\textsc{1.5B}\xspace}

\NewDocumentCommand{\gptFiveTwo}{}{\textsc{Gpt}-\textsc{5.2}\xspace}
\NewDocumentCommand{\geminiThreePro}{}{\textsc{Gemini}-3-\textsc{Pro}\xspace}

\RequirePackage[
 datamodel=acmdatamodel,
 style=acmnumeric, 
 uniquelist=false,
 backend=biber,
 maxcitenames=1,
 mincitenames=1,
 minbibnames=1,
 maxbibnames=1
]{biblatex}

\input{includes/biblatex_shim}

\begin{document}


\title{How Robustly do LLMs Understand Execution Semantics?}

\author{Claudio Spiess}
\email{cvspiess@ucdavis.edu}
\affiliation{%
  \institution{UC Davis}
  \city{Davis}
  \state{California}
  \country{USA}
}

\author{Prem Devanbu}
\email{pdevanbu@ucdavis.edu} 
\affiliation{%
  \institution{UC Davis}
  \city{Davis}
  \state{California}
  \country{USA}
}

\author{Earl T. Barr}
\email{e.barr@ucl.ac.uk} 
\affiliation{%
  \institution{University College London}
  \city{London}
  \country{UK}
}

\renewcommand{\shortauthors}{Spiess et al.}

\begin{abstract}
LLMs demonstrate remarkable reasoning capabilities, yet whether they utilize internal world models or rely on sophisticated pattern matching remains open. 
We study LLMs through the lens of \emph{robustness of their code understanding} using a standard program-output prediction task. 
Our results reveal a stark divergence in model behavior: while open-source reasoning models (DeepSeek-R1 family) maintain stable, albeit somewhat lower accuracies (\qty[round-mode=figures, round-precision=2]{38.3}{\percent} to \qty[round-mode=figures, round-precision=2]{67.11}{\percent}) under code transformations
\& input perturbations, the frontier model GPT-5.2 exhibits significant brittleness. 
Despite achieving a near-perfect score of \qty[round-mode=figures, round-precision=2]{99}{\percent}
on the original, unperturbed \cruxeval benchmark, perturbed inputs trigger accuracy declines between \qty[round-mode=figures, round-precision=2]{19.69}{\percent} and \qty[round-mode=figures, round-precision=2]{23.63}{\percent}. In addition, we find that many models perform
much worse at predicting behavior on perturbed inputs that raise exceptions, \emph{and} that
prediction performance depends on the kind of exception. We study
remedies to address this deficiency in exception prediction, and evaluate the effect of these
remedies on the ability to predict non-exception behaviors. 
Our findings both point to limitations in the way all models understand code, and establish the value of using perturbation to evaluate code models.
\end{abstract}

\ccsdesc[500]{Computing methodologies~Artificial intelligence}

\keywords{code understanding, robustness, metamorphic testing}


\maketitle

\section{Introduction}
\label{sec:intro}
Current LLMs show impressive proficiency in 
Software Engineering tasks, such as code generation~\cite{jiangSurveyLargeLanguage2025}, code summarization~\cite{sunSourceCodeSummarization2025}, and
code repair~\cite{yangSurveyLLMbasedAutomated2025}. This  raises
the question: to what extent does language  model
performance arise from actually 
\emph{understanding} the semantics of code? 
Studies show that developers 
spend the majority of their time understanding existing
code~\cite{xia2017measuring, kemerer1995software}; the capacity for
\emph{understanding} code is thus
vital to human coding work. 
For example, maintenance work can require understanding what
a buggy program actually does, in response to an failure-triggering input. Thus, it is natural to ask if LLMs are similarly able to understand~code. 

Language model ``understanding'' is a topic well-explored for natural language, using benchmarks
such as MMLU~\cite{hendrycksMeasuringMassiveMultitask2020} and SuperGLUE~\cite{wangSuperGLUEStickierBenchmark2019}. 
Recently,
several benchmarks have tackled this issue
for code. CodeMMLU~\cite{nguyenCodeMMLUMultiTaskBenchmark2025} comprises multiple-choice
tests that
evaluate several types of code understanding (syntax
and semantics). \cruxeval~\cite{guCRUXEvalBenchmarkCode2024}
is more demanding, requiring LLMs to understand
programs well-enough to \emph{precisely
predict exact outputs} for given test inputs (not
just select one answer from presented choices,~as~with~CodeMMLU).

To effectively maintain a program, 
one must understand it sufficiently well to \emph{robustly} and consistently predict its
behavior, regardless of how exactly it
is coded, which inputs it is given, or the
intricacy of their execution paths. Predicting  outputs given inputs is a vital capability for software maintenance tasks: developers
need to do this when diagnosing failures,
when refactoring the code or adding enhancements.
LLMs perform surprisingly well on this task, with frontier models predicting outputs for inputs for published benchmarks~\cite{guCRUXEvalBenchmarkCode2024} with over 99\% accuracy, but
how \emph{{\bf \ul{robust}}} is this performance? 
In other words, how much can a developer rely on LLM predictions while debugging or maintaining~code?

Prior work suggests that language model understanding
of code is \emph{not} robust with respect
to program structure for highly demanding tasks
like code equivalence and static analysis~\cite{hoodaLargeCodeModels2024,weiEquiBenchBenchmarkingLarge2025}. In this paper, we aim to refine the benchmarks for evaluating LLM code understanding by treating output prediction as a proxy for code understanding. We define robustness as the ability of a model to consistently correctly predict a program's output on a given input despite variations that do not alter the program's semantics. To measure it, our study contributes an evaluation of LLM robustness across three critical dimensions:

First, we evaluate the robustness of LLM ability to predict program outputs, \emph{\ul{when subject to input perturbation}}. 
We apply type-aware mutation to generate a dense neighborhood of local inputs for each program; we find that LLMs frequently do not make consistent (all right or all wrong) output predictions,
even within these constrained input spaces, indicating a weak understanding of the code. 
Some inputs cause a program to fail; 
the LLM should be able to correctly and consistently predict behavior for such failing~runs~as~well. 

\rqs{How robust is LLM code understanding to input 
perturbation? How well do LLMs predict runtime exceptions for perturbed inputs that cause exceptions?}

\noindent\emph{Finding Summary: }We find evidence of lack of robustness, both (unexpectedly) in some large, frontier models, 
\emph{and} also with exception-raising inputs. We explore these issues in detail.

Next, we consider robustness of LLMs on the output prediction task, 
\emph{\ul{when subject to program transformation}}:
we use meaning-preserving transformations (MPTs) 
to evaluate LLM performance on output prediction by comparing a model's performance on syntactically different, but semantically equivalent, variants of an~initial~program.

\rqs{How robust is LLM code understanding to program perturbation via meaning-preserving code transformations?}

\noindent\emph{Finding Summary: }
With a few unexpected exceptions, notably in frontier models, we find that models mostly 
respond robustly to the meaning-preserving transformations in our study.

Finally, we study the robustness of LLMs, on the same output prediction task, 
\emph{\ul{when execution traces have varying numbers of decisions}}:
we use dynamic analysis (tracing) to evaluate LLM performance on output prediction 
as a function of the number of decisions encountered during execution.

\rqs{How robust is LLM code understanding to control flow decisions? Are execution paths with more decisions, more difficult to reason about?}

\noindent\emph{Finding Summary: }
In general, our data broadly suggests that LLMs are less
accurate at predicting outputs as more decisions are encountered along an execution trace.
When the decisions are primarily loop control conditions, as opposed to if-else chains, this finding points to disfluencies in LLM's understanding of iteration.

\section{Background}
\label{sec:background}
To evaluate the robustness of LLMs on code execution tasks, we distinguish between standard performance and structural consistency. Given a set of programs \(\mathcal{P}\), where each program \(p \in \mathcal{P}\) is associated with a canonical input \(x_{p,0}\) (from the original \cruxeval benchmark) and a set of \(n\) synthetic variants \(\{x_{p,1}, \dots, x_{p,n}\}\), we define the following metrics:

\subsection{Performance Metrics}
\begin{itemize}
    \item \textbf{Accuracy \(C_{o}\)}: The model's accuracy on the canonical \cruxeval input set. It serves as the baseline for comparison with prior work, and is equivalent to Pass@1.
    \item \textbf{Accuracy \(C_{\tilde{o}}\)}: The model's average accuracy across all perturbed input variants \cruxevalO, excluding the original.
    \item \textbf{Accuracy \(C_{o \, \cup \, \tilde{o}}\)}: The mean accuracy across the union of original and perturbed inputs.
\end{itemize}

\subsection{Robustness and Generalization}
\begin{itemize}
    \item \textbf{Program-level Strict Robustness (PSR)}: We introduce this new robustness measure. 
 It is the fraction of programs on which a model predicts the correct output for \emph{all} \(n\) inputs; PSR measures how likely a model is to predict output correctly for any input to an arbitrary program.  

    \begin{equation}
        \text{PSR} = \frac{1}{|\mathcal{P}|} \sum_{p \in \mathcal{P}} \prod_{i=1}^{n} \mathds{1}(f(p, x_{p,i}) = y_{p,i})
    \end{equation}

    \item \textbf{Robust Drop (\rdel)}: The absolute performance degradation when moving from canonical to perturbed inputs. The greater the magnitude of the drop, the less robust the model's performance. 
    \begin{equation}
        \rdel = \text{Accuracy}_{C_{\tilde{o}}} - \text{Accuracy}_{C_{o}}
    \end{equation}
\end{itemize}
\section{Methodology}
\label{sec:methodology}

This section presents our methodology, starting with our experimental design for each RQ, and closing with our model selection.

\subsection{Experimental Design}
\label{sec:expdesign}

We now detail our experimental design for each of our RQ.

\noindent\paragraph{RQ1}{
\label{para:rq1}We perturb \cruxeval inputs using a type-aware mutation algorithm following \textcite{liuYourCodeGenerated2023}. We produce perturbed inputs using the original input \& program pairs in \cruxeval, using rejection sampling. We mutate the original \cruxeval input, reject the input if it is already in the set of inputs for that particular program, and otherwise execute the code with that input, record the output, and 
save samples that raise exceptions separately in
a new \cruxevalExc dataset. If an exception is raised, we record its type and message \eg \texttt{IndexError} and `\texttt{list index out of range}'. 

As \cruxeval programs are small and simple, we enforce a 5 second execution time limit, which when exceeded raises a custom \texttt{TimeoutError} exception, suggesting an infinite loop or recursion\footnote{\textcite{guCRUXEvalBenchmarkCode2024} use a 3 second timeout, which we increase to 5 due to local compute resource constraints.}. We limit the number of distinct perturbed inputs to 10 due to experimental (LLM usage) costs. Some programs have fewer than 10 inputs due to implicit input constraints \eg a program may enforce acceptable values for some argument \(x\) with a cardinality \(< 10\). We exclude such programs from our study to ensure equal sample sizes, and leave constraint-aware perturbation to future work.

The ability to predict that an exception will be raised while executing a given program-input pair \((p, x_{p,i})\), and specifically \emph{which} exception, is helpful while debugging \& maintaining code. \cruxeval ignores
this issue; we, however, also gather \& include  input samples that raise runtime exceptions in our enhanced \cruxevalExc, thus extending output prediction to cover exceptions.

For exception prediction, we initially labeled a response as correct if the ground-truth exception type or message was present anywhere in the model generation. Upon reviewing model generations, we found exceptions that were correctly predicted, but did not contain the exception type verbatim nor the exact message \eg for ValueError: `\texttt{This code will crash because arguments must have the same length}'. To minimize false negatives, we subsequently devised heuristics for TimeoutError, ValueError, TypeError, and IndexError. For each, we define 1-5 substrings that indicate a correct prediction \eg `\texttt{infinite}' for TimeoutError.

We utilize two notions of correctness: \begin{enumerate*}[label=\arabic*)]\item Any Match: \emph{any} exception \emph{type} or any heuristic substring is present in the model's response, regardless of what exception is recorded during execution (ground truth). This captures the ability to predict that \emph{some} exception will occur. \item Strict Match: The ground-truth exception type or message is present in the model's response, or a heuristic substring is present for that \emph{particular} exception. In contrast to 1), this captures the ability to predict \emph{which} exception will occur.\end{enumerate*}

Our evaluation begins with the original \cruxeval few-shot prompt~\textcite{guCRUXEvalBenchmarkCode2024}, as shown in~\autoref{fig:prompts}. 
The answers are in the form 
{\small\ttfamily [ANSWER] assert \$expr\$ == \$value\$ [/ANSWER]}.
We extract the predicted result {\small\ttfamily \$value\$} from model responses using heuristics and regular expressions, ensuring that left-hand side of the assertion matches the query ground truth. We compare the extracted output (right-hand side) with the gold-truth output recorded during execution using a Python expression evaluator\footnote{\url{https://pypi.org/project/asteval/}}, minimizing the effects of stylistic differences such as quotations and whitespace. We found that in some cases, models produced explanations, chain-of-thought, or further examples using the provided function, subsequent to the correct answer. To not penalize models in such cases, we label a response containing multiple assertions as correct, if \emph{any} of the extracted assertions are correct, \ie both the left and right-hand side of the assertion match the ground truth.

We study multiple open-source models and three frontier models. We include base models, instruction-tuned models, and both models trained for general reasoning, and code reasoning.

\noindent\paragraph{RQ2}{

We perturb the original \cruxevalO programs with meaning-pre\-serving transformations (MPTs), producing \cruxevalMpt with four groups detailed below as \emph{Syntax Transform}. We additionally perturb \cruxevalO with variable renaming to produce \cruxevalRenamed, further detailed below as \emph{Variable Renaming}. We execute the code with the original input and verify output equivalence as a sanity-check of semantic preservation.

We hypothesize that the combination of MPTs and input perturbations would create an antagonistic effect, further degrading LLM performance on the \cruxeval output prediction task. To help characterize these dynamics, we isolate the individual contributions of each class of perturbations in this paper. This granular approach is a prerequisite for understanding the underlying failure modes before addressing the compounding complexities of their interactions, which we reserve for future inquiry.

\textbf{Syntax Transform} Following \textcite{chakrabortyNatGenGenerativePretraining2022}, we apply OperandSwap, BlockSwap, ForWhileSwap, and DeadCodeInsertion transformations, producing \cruxevalMpt.

\textbf{Variable Renaming} We mine \textsc{DyPyBench}, a dataset of 50 large, popular Python projects~\cite{bouzeniaDyPyBenchBenchmarkExecutable2024}, for identifiers and their occurrences. For each \cruxeval program, we rename all identifiers with randomly sampled ones, weighted by number of occurrences, from the mined identifier distribution, \ie the most common identifiers are more likely to be sampled. This ensures that the identifiers are unlikely to be rare, and thus carry higher likelihood for the model, while also likely to be out-of-context. Thus, the renamed variables are generic, and unlikely to be relevant; we refer to this setting~as~\cruxevalRenamed.

We repeat our model evaluation as described in~\autoref{sec:expdesign}.

}

\paragraph{RQ3}{We employ both static analysis to identify programs with loop constructs, and dynamic analysis (\emph{tracing}) to record execution behavior for a given input \eg executed line-numbers, number of loop iterations, \etc We focus on loops, \ie \pymint{for}/\pymint{while} and comprehensions. We record the number of decisions encountered in each loop \eg a \pymint{for} loop with one \pymint{if}, with a compound condition \eg \pymint{(x > 5) and (y < 3)} in the header (and no further conditions in its body), would have a total of \textbf{three} decisions: the iterator itself, and the two sub-expressions in the \pymint{if} condition.
With two iterations, the execution
encounters a total of \emph{six} decisions.
}

\subsection{Model Selection}
To provide a view of how LLM architecture influences code understanding, we selected a diverse suite of 14 models. Our selection criteria prioritized model size, training regime, and access model (open-weights \vs proprietary). The resulting cohort includes:
Qwen2.5 Math 1.5B and 7B~\cite{yangQwen25MathTechnicalReport2024},
DeepSeek R1 Distilled 1.5B, 7B, 8B~\cite{deepseekr12025}, 
Qwen2.5 Base and Instruct (7B and 14B)~\cite{qwenQwen25TechnicalReport2025},
NVIDIA OpenCodeReasoning Nemotron 7B and 14B~\cite{ahmadOpenCodeReasoningAdvancingData2025},
Llama 3.1 8B~\cite{grattafioriLlama3Herd2024},
GPT-5 Nano,
GPT-5.2, and
Gemini 3 Pro. We cover a parameter size range from 1.5B to 14B (proprietary model size unknown at time of writing), training regimes completion, instruction-following, reasoning, code-reasoning, and access model (3 proprietary, 11 open-weights). We also note that each of the DeepSeek distilled models has its non-reasoning ancestor.

To follow developer workflows using frontier models, which often neglect manual parameter adjustment~\cite{donatoStudyingHowConfigurations2025}, we accessed GPT-5.2, GPT-5 Nano, and Gemini via their default API configurations. To mitigate experimental bias, we performed each query independently to avoid context-leakage~\cite{guAuditingPromptCaching2025}.
No explicit system prompts were utilized, ensuring the evaluation focused strictly on the models' baseline performance and inherent knowledge~\cite{muCloserLookSystem2024}.
\section{Results}
\label{sec:results}
We evaluate model performance across our three research questions, measuring robustness to input perturbation (RQ1), code perturbation (RQ2), and execution decisions (RQ3). The following sections report the findings of our experiments across our studied models.

\subsection{RQ1}
\label{sec:rq1-results}
\begin{table}[b]
\centering
\caption{Accuracy \& robustness metrics for original \vs perturbed \cruxeval \emph{inputs}. Arrow \textuparrow{} indicates higher values are preferred. Each model was evaluated on 684 programs, for which the original benchmark input and 10 unique perturbed variants were available.}
\label{tab:rq1pinputresults}
\begin{tblr}{
  colspec = {X[l,2] X[r] X[r] X[r] X[r] X[r]},
  row{1} = {font=\bfseries,halign=c},
  hline{1,2,Z} = {0.8pt},
  hline{3-Y} = {dashed},
  columns = {font=\small},
  stretch = 0.8,
  rowsep = 0.5pt,
  colsep = 1pt
}
Model & 
\(C_{o}\) (\textuparrow) & 
\(C_{\tilde{o}}\) (\textuparrow) & 
\(C_{o \, \cup \, \tilde{o}}\) (\textuparrow) & 
PSR (\textuparrow) & 
\rdel (\textuparrow) \\
Qwen2.5 Math 1.5B & \num{12.481644640234949} & \num{12.21732745961821} & \num{12.24135629421973} & \num{1.908957415565345} & \num{-0.26431718061673964} \\
DS R1 1.5B & \num{38.45029239766082} & \num{37.5} & \num{37.58639021796917} & \num{4.385964912280701} & \num{-0.950292397660818} \\
Qwen2.5 Math 7B & \num{34.941520467836256} & \num{33.2748538011696} & \num{33.42636895268474} & \num{11.11111111111111} & \num{-1.6666666666666607} \\
Qwen2.5 Instr 7B & \num{40.93567251461988} & \num{41.812865497076025} & \num{41.73312068048911} & \num{16.812865497076025} & \num{+0.8771929824561431} \\
Nemotron 7B & \num{46.198830409356724} & \num{43.260233918128655} & \num{43.52737905369484} & \num{4.385964912280701} & \num{-2.9385964912280684} \\
DS R1 7B & \num{65.05847953216374} & \num{62.61695906432748} & \num{62.83891547049441} & \num{27.339181286549707} & \num{-2.4415204678362556} \\
Llama 3.1 8B & \num{36.40350877192983} & \num{33.81578947368421} & \num{34.05103668261563} & \num{11.842105263157894} & \num{-2.5877192982456165} \\
DS R1 8B & \num{64.91228070175438} & \num{62.909356725146196} & \num{63.09144072301967} & \num{29.678362573099413} & \num{-2.002923976608195} \\
Qwen2.5 14B & \num{22.076023391812864} & \num{23.20175438596491} & \num{23.099415204678362} & \num{3.8011695906432745} & \num{+1.1257309941520437} \\
Qwen2.5 Instr 14B & \num{47.214076246334315} & \num{47.84457478005864} & \num{47.78725673153825} & \num{21.114369501466275} & \num{+0.630498533724333} \\
Nemotron 14B & \num{67.2514619883041} & \num{65.05847953216374} & \num{65.25784157363105} & \num{33.47953216374269} & \num{-2.1929824561403577} \\
GPT-5 Nano & \num{96.7836} & \num{96.6082} & \num{96.6241} & \num{82.7485} & \num{-0.175439} \\
GPT-5.2 & \SetCell{font=\small\bfseries} \num{99.12280701754386} & \num{79.60526315789474} & \num{81.37958532695376} & \num{56.14035087719298} & \num{-19.517543859649123} \\
Gemini 3 Pro & \num{99.1044776119403} & \SetCell{font=\small\bfseries} \num{99.35820895522389} & \SetCell{font=\small\bfseries} \num{99.33514246947081} & \SetCell{font=\small\bfseries} \num{97.16417910447761} & \SetCell{font=\small\bfseries} \num{+0.2537313432835919} \\
\end{tblr}
\vspace*{-1\baselineskip}
\end{table}

In~\autoref{tab:rq1pinputresults}, we observe that accuracy ranges widely across models, from as little as \pct{12.48}, to as high as \pct{99.12}. Our results show the effectiveness of reasoning post-training \eg \qwenMathOneB's accuracy jumps from \pct{12.48} to \pct{38.45} after the DeepSeek R1 distilled finetune. Our best open-source model reaches \pct{67.25} accuracy, far lower than the best-performing frontier model GPT-5.2's accuracy of \pct{99.12}.

Considering \rdel, we observe four models that achieved higher accuracy on the perturbed inputs than the original input, resulting in positive \rdel values. On the other hand, the majority of open-weight models did worse on the perturbed inputs, with an average \rdel of \numrnd[3]{-1.1272727273}. In contrast, the frontier model \gptFiveTwo suffered a significant \rdel of \numrnd[2]{-19.52}.

We also report (PSR) numbers, the proportion of programs for which models predict \emph{all outputs
correctly}. These are much lower, reaching only  \pct[0]{56.14035087719298} for \gptFiveTwo. Arguably,
correct performance for \emph{several inputs for the same program} is a better measure of how well a model actually understands a program; if a model truly understands a program, it should get the output correct on all inputs! The drop for \gptFiveTwo from \pct[0]{99} overall correct \emph{for all inputs}
to just \pct[0]{56} correct \emph{for all programs},
suggests that 
its output prediction capability for a random program on \emph{any} given input is much less reliable than its \pct[0]{99} performance on the original \cruxeval might indicate.
We explore this concerning
issue further in~\autoref{sec:discussion}.

A surprising performance gap exists between GPT-5 Nano and GPT-5.2. While GPT-5 Nano is marketed as a smaller, more efficient version of GPT-5 released in August 2025, and GPT-5.2 is marketed as a flagship model released in December 2025, GPT-5 Nano outperforms GPT-5.2 on \(C_{\tilde{o}}\) by \pct[0]{17}, \(C_{o \, \cup \, \tilde{o}}\) by \pct[2]{15.24}, and PSR by \pct[2]{26.61}. Furthermore, its \rdel of \num{-0.18} is greater than the \rdel of \num{-19.52} for GPT-5.2. These results suggest that GPT-5 Nano is more robust than its nominally stronger counterpart. We hypothesize that it is a quantized, and possibly distilled, version of GPT-5. Prior work~\cite{askarihemmatQRegRegularizationEffects2022,javedQTDoGQuantizationAwareTraining2025,bouguerraLessPreciseCan2026} suggests quantization exerts a regularizing effect, steering optimization toward flatter minima
that exhibit greater robustness to perturbation, and a reduced tendency for overfitting. If our hypothesis holds, quantization-regularization may help explain the greater generalization we observe for GPT-5 Nano.

Notably, Gemini 3 Pro exhibits very high performance and impressive robustness: its \rdel is small and \emph{positive}, and its PSR on the perturbed datasets is not that much lower than its performance on original \cruxeval dataset.

\begin{table}[b]
\centering
\caption{Performance of the frontier models on accuracy of prediction
the precise exception that actually occurs (``Strict Match'') or that
some exception occurs (``Any Match'')}
\label{tab:exceptionsOriginal}
\begin{tblr}{
  colspec = {X[l] X[r] X[r]},
  row{1} = {font=\bfseries,halign=c},
  hline{1,2,Z} = {0.8pt},
  hline{3-Y} = {dashed},
  columns = {font=\small},
  stretch = 0.8,
  rowsep = 0.5pt,
}
  Model & Strict Match \% & Any Match \% \\
  GPT-5 Nano & \SetCell{font=\small\bfseries} \num{71.9468} & \SetCell{font=\small\bfseries} \num{73.156} \\
  GPT-5.2 & \num{14.7659} & \num{14.7659} \\
  Gemini 3 Pro & \num{48.5612} & \num{50.7194} \\
\end{tblr}
\end{table}

We subsequently investigated, for the best-performing frontier models, the LLM's ability to reason about input perturbations that result in runtime exceptions. 
In~\autoref{tab:exceptionsOriginal}, we report the frontier model performances on \cruxevalExc using the canonical \cruxeval prompt, for both correctness criteria.
In general, we observe that 
models find it more difficult to predict exceptions. \autoref{tab:exceptionsOriginal} shows the accuracy of our 3 best-performing
frontier models on predicting exceptions; the first column is a strict match, 
evaluating if the model can predict precisely which exception is thrown; the
second column is ``Any'' which credits the model when any exception is predicted. 
It's noteworthy that models actually manage to predict the \emph{right} exception in most cases that they manage to correctly predict that \emph{some} exception is thrown. 
In~\autoref{sec:discussion}, we explore the type of exceptions that are strictly correctly
predicted, and then delve into the
factors 
potentially related to poor exception performance, including the potentially prompt-compliant (``sychophantic'') tendency of instruction-tuned LLMs to produce a plausible, fake, non-erroneous output, even when the actual output is an error.

\begin{answerenv}{1}
We find that most models perform worse with the perturbed inputs (average
\rdel of \numrnd[1]{-1.13}), with \gptFiveTwo being the most affected (\rdel of \numrnd[1]{-19.52}); furthermore, we notice a big performance degrade on 
predicting exceptions; we explore this further in~\autoref{sec:discussion}.
\end{answerenv}

\subsection{RQ2}
\label{sec:rq2Results}
In~\autoref{tab:model_results_rq2}, we observe generally minor differences when shifting from the original \cruxeval programs to their transformed (with Meaning-Preserving Transformation, ``MPT'') and Renamed variants. In contrast to~\autoref{sec:rq1-results}, we only perturb the program \emph{code} in this setting, as opposed to the \emph{inputs} \ie test cases.

\begin{table}[b]
\centering
\caption{Average model accuracy on canonical \cruxeval, code-perturbed \cruxevalMpt, and \cruxevalRenamed. Asterisks *, **, and *** represent McNemar test \(p < 0.05\), \(0.01\), and \(0.001\) respectively, for original \vs code-perturbed. As each model is a different experiment, we do not make claims across all models based the \pvalues and thus do not apply corrections.}
\label{tab:model_results_rq2}
\begin{tblr}{
  colspec = {X[l,2] X[r] X[r] X[r]},
  row{1} = {font=\bfseries,halign=c},
  hline{1,2,Z} = {0.8pt},
  hline{3-Y} = {dashed},
  columns = {font=\small},
  stretch = 0.8,
  rowsep = 0.5pt,
}
Model & \cruxeval & \cruxevalMpt & \cruxevalRenamed \\
Qwen2.5 Math 1.5B & \num{12.25} & \num{9.26966}* & \num{11.125} \\
  DS R1 1.5B & \num{40} & \num{40.4494} & \num{35.5}* \\
  Qwen2.5 Math 7B & \num{37} & \num{35.5337} & \num{34.375} \\
  Qwen2.5 Instr 7B & \num{42.625} & \num{42.9775} & \num{40.5} \\
  Nemotron 7B & \num{47} & \num{37.2191}*** & \num{44.75} \\
  DS R1 7B & \num{62.625} & \num{60.2528} & \num{61.5} \\
  Llama 3.1 8B & \num{37.875} & \num{34.5506} & \num{34.625}* \\
  DS R1 8B & \num{66.125} & \num{61.6573} & \num{61.25}** \\
  Qwen2.5 14B & \num{22.5} & \num{27.5281} & \num{16}*** \\
  Qwen2.5 Instr 14B & \num{48.875} & \num{45.6461} & \num{47} \\
  Nemotron 14B & \num{68.625} & \num{57.8652}*** & \num{66.125} \\
  GPT-5 Nano & \num{97.2046} & \num{87.9213}*** & \num{93.875}*** \\
  GPT-5.2 & \SetCell{font=\small\bfseries} \num{99.2376} & \num{76.4045}*** & \num{75.75}*** \\
  Gemini 3 Pro & \SetCell{font=\small\bfseries}  \num{99.2376} & \SetCell{font=\small\bfseries} \num{95.7865}*** & \SetCell{font=\small\bfseries} \num{98.3689} \\
\end{tblr}
\end{table}

Similar to our observations in~\autoref{tab:rq1pinputresults}, the reasoning distilled models show remarkable gains over their base counterparts. For instance, Llama 3.1 8B accuracy on \cruxeval jumps from \pct{37.23} to \pct{67.82} in its DeepSeek R1 Distill (``DS R1 8B'' in~\autoref{tab:model_results_rq2}) version. However, these models also degrade under perturbation, with DeepSeek R1 Distill 8B dropping to \pct{61.66} on MPT (\numrnd[2]{-6.16}~\(\Delta\)) and \pct{61.25} (\numrnd[2]{-6.57}~\(\Delta\)) on Renamed. Among these open-source models, DeepSeek R1 Distill 14B emerges as the strongest performer, achieving \pct{76.75} on the original programs, and maintaining high scores across both perturbation settings. Unexpectedly, the 14B model outperforms its 32B variant. We observe a surprisingly narrow performance gap between R1 14B and GPT-5.2. 

The impact of variable renaming and MPTs is particularly evident for the frontier model GPT-5.2, showing a sharp decline from a near-perfect \pct{99.24} to around \pct{76.40} 
for both MPT and Renamed. 
In contrast, the decline for Gemini 3 Pro is much smaller, dropping from \pct{99.24} to \pct{95.79} and \pct{98.37} for MPT and Renamed, respectively. 
On our dataset, despite matching \geminiThreePro's performance on the original \cruxeval, \gptFiveTwo is more sensitive to surface-level syntactic code changes than \geminiThreePro.
This is surprising as GPT-5.2 is near-perfect on the original benchmark, is newer (Jan 2025 \vs Dec 2025), and likely larger in terms of parameter count. GPT-5 Nano, while also being a strong performer on the original
code, is significantly impacted by MPT and Renaming although not as much as GPT-5.2 \emph{per se}. 

\begin{answerenv}{2}
We find that in most cases, the performance decreases somewhat
for \cruxevalMpt, and even increases in a couple of cases (DeepSeek R1 1.5B, and Qwen 2.5 14B). For \cruxevalRenamed (variable renaming), performance does consistently drop (average \rdel of \numrnd[1]{-4.316851}), but the differences are surprisingly low. 
The most noteworthy change is that one of the best performing models, GPT-5.2, suffers a dramatic performance drop (\(\approx\)\numrnd[0]{-23.160364}) on both renaming and transformation; the other frontier model, however, remains more robust. 
\end{answerenv}

\subsection{RQ3}
We examine the results of decisions encountered during execution.
\begin{figure}[b]
	\centering
    \includegraphics[width=\columnwidth]{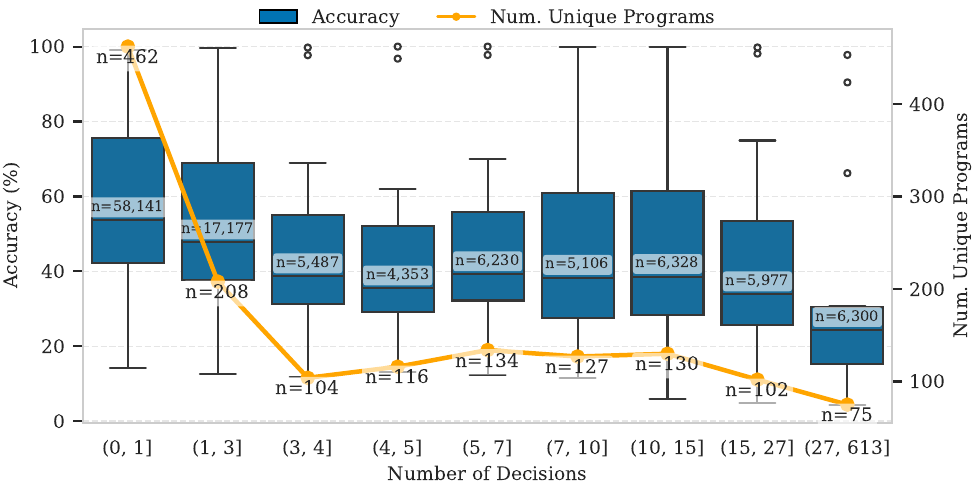}
	\caption{Number of decisions encountered on execution path \vs model accuracy on output prediction. The yellow line shows the varying number of sample programs with increasing numbers of conditions on the execution paths}
	\label{fig:fig2}
\end{figure}
In~\autoref{fig:fig2}, we visualize the results from~\autoref{tab:rq3result}. We investigate the relationship between the number of decisions encountered on an execution path for a given program-input pair \((p, x_{p,i})\), and model accuracy. Each box in the figure represents the 14 model accuracies across \(n\) samples of output predictions for program-input pairs.

Due to the long-tail distribution of decision counts that arises from certain program-input pairs having many loop iterations, we bin the decision counts into nine bins for sufficient sample sizes---one for no decisions encountered (\eg straight line code), and eight \emph{quantile} bins, with roughly comparable sample sizes varying from \intnum{4353} to \intnum{17177} observations. 

Additionally, we plot the number of unique programs in each bin on the second (right) \(y\)-axis. Given that the sum of unique programs (\intnum{1458}) exceeds the number of programs in \cruxeval, it follows that execution paths and thus decision counts vary between inputs for the same program.

\begin{table}[t]
\centering
\caption{Binned number of decisions encountered during execution \vs mean model accuracy. Columns report number of samples (model, program, input triples), unique programs and unique inputs per bin.}
\label{tab:rq3result}
\begin{tblr}{
  colspec = {X[r] X[r] X[r] X[r] X[r]},
  row{1} = {font=\bfseries,halign=c},
  hline{1,2,Z} = {0.8pt},
  hline{3-Y} = {dashed},
  columns = {font=\small},
  stretch = 0.8,
  rowsep = 0.5pt,
}
  Number of Decisions &  Samples & Accuracy (\%) & Unique Programs & Unique Inputs \\
  (0, 1] & \intnum{58141} & \num{58.5353} & \intnum{462} & \intnum{4154} \\
  (1, 3] & \intnum{17177} & \num{55.4753} & \intnum{208} & \intnum{1227} \\
  (3, 4] & \intnum{5487} & \num{47.4941} & \intnum{104} & \intnum{392} \\
  (4, 5] & \intnum{4353} & \num{45.1643} & \intnum{116} & \intnum{311} \\
  (5, 7] & \intnum{6230} & \num{48.0578} & \intnum{134} & \intnum{445} \\
  (7, 10] & \intnum{5106} & \num{48.4528} & \intnum{127} & \intnum{365} \\
  (10, 15] & \intnum{6328} & \num{48.4197} & \intnum{130} & \intnum{452} \\
  (15, 27] & \intnum{5977} & \num{44.7549} & \intnum{102} & \intnum{427} \\
  (27, 613] & \intnum{6300} & \num{33.6508} & \intnum{75} & \intnum{450} \\
\end{tblr}
\vspace*{-1\baselineskip}
\end{table}

We observe a negative relationship between number of decisions and model output prediction correctness.
A Mann-Whitney \(U\) test on pooled model results indicated that the number of decisions encountered was significantly lower for correct predictions than for incorrect ones (\(U = \num[round-precision = 3]{1.436e9}\), \(p<.001\)). While the effect size as measured by the rank-biserial correlation was weak (\(r_{rb} = \num[round-precision = 3]{-.129}\)), this  
suggests that, as the number of decisions a model must reason about increases \eg an \pymint{if} statement in a \pymint{for} loop, its ability to correctly reason about the code execution decreases. Further analysis per model found a minimum \(r_{rb}\) of \num[round-precision = 3]{-0.314} (Nemotron 14B), a median of \num[round-precision = 3]{-0.167}, and maximum of \num[round-precision = 3]{0.273} (Gemini 3 Pro). As the only model with a positive (surprisingly strong) effect, Gemini may be less hindered by the number of decisions.

Despite the overall negative trend in~\autoref{fig:fig2}, we note an increase in accuracy in the (5, 15] interval, where accuracy seems to recover somewhat from that in (3, 5]. The samples spike at this interval;  further examination of this phenomenon and the Gemini 3 Pro outlier is left to future work.

\begin{answerenv}{3}
We find that as the number of decisions increases to 3, there is a significant drop off in prediction accuracy; models generally struggle to predict outcomes as the number decisions encountered increases; however, our data suggests that Gemini 3 Pro's prediction accuracy in particular, does not decrease as decisions encountered increases.
\end{answerenv}
\section{Discussion}
\label{sec:discussion}
We begin by noting the dramatically poor performance of frontier models in predicting outputs for \cruxevalExc compared to the non-exception-raising input set \eg GPT-5.2 accuracy drops from \(\sim\)81\% in~\autoref{tab:rq1pinputresults}, to
\(\sim\)15\% for exception prediction in~\autoref{tab:exceptionsOriginal}
(also reproduced in the ``Original'' line in~\autoref{tab:exceptions}).

\autoref{tab:exceptionscounts} breaks down the accuracy of the models in predicting exceptions, based on the type of exception actually
raised when running the code sample.
Due to the small sample sizes for several exception types in \cruxevalExc, the zero and \pct[0]{100} accuracy results in~\autoref{tab:exceptionscounts} lack a reliable basis for comparison and are therefore excluded from this analysis.

The most commonly thrown exception is TypeError, with 358 occurrences; the least common is NameError, with just 3
occurrences. Each exception is a standard Python exception, with the exception of TimeoutError, which occurs when the code took more than 5 seconds (wall time) to execute (see~\sref{para:rq1}), suggesting an infinite loop. We manually confirmed that the 11 programs for which this occurred were indeed infinite \pymint{while} loops.
\begin{table}[b]
\centering
\caption{Distribution of exception types in \cruxevalExc, along with model performance using the original \cruxeval prompt.}
\label{tab:exceptionscounts}
\begin{talltblr}[note{a} = {GPT-5 Nano},
  note{b} = {GPT-5.2},
  note{c} = {Gemini 3 Pro}]{
  colspec = {X[l, 2] X[r] *{3}{X[r] X[r]}},
  row{1,2} = {font=\bfseries,halign=c},
  row{2} = {font=\tiny},
  hline{1,3,Z} = {0.8pt},
  rowspec={Q|[gray]},
  hline{4-Y} = {dashed},
  columns = {font=\small},
  stretch = 0.8,
  rowsep = 1pt,
  vline{3} = {2-Z}{dotted},
  vline{6} = {2-Z}{dotted},
  cell{1}{3}={c=3}{c},
  cell{1}{6}={c=3}{c},
  cell{1}{1,2}={r=2}{c}
}
   Exception & Count & Any Match \% (\textuparrow) & & & Strict Match \% (\textuparrow) \\
 & & Nano\TblrNote{a} & 5.2\TblrNote{b} & Gem3\TblrNote{c} & Nano\TblrNote{a} & 5.2\TblrNote{b} & Gem3\TblrNote{c} \\
  TypeError & \intnum{358} & \num{58.8235} & \num{5.32213} & \num{31.5642} & \num{56.0224} & \num{5.32213} & \num{27.6536} \\
  ValueError & \intnum{152} & \num{81.5789} & \num{15.1316} & \num{75.6579} & \num{81.5789} & \num{15.1316} & \num{75.6579} \\
  IndexError & \intnum{138} & \num{95.4887} & \num{24.6377} & \num{62.3188} & \num{95.4887} & \num{24.6377} & \num{61.5942} \\
  Attribute\-Error & \intnum{73} & \num{94.4444} & \num{1.36986} & \num{26.0274} & \num{94.4444} & \num{1.36986} & \num{26.0274} \\
  KeyError & \intnum{54} & \num{94.4444} & \num{83.3333} & \num{98.1481} & \num{94.4444} & \num{83.3333} & \num{98.1481} \\
  Timeout\-Error & \intnum{36} & \num{27.7778} & \num{0} & \num{63.8889} & \num{27.7778} & \num{0} & \num{63.8889} \\
  Recursion\-Error & \intnum{10} & \num{100} & \num{0} & \num{100} & \num{100} & \num{0} & \num{70} \\
  Lookup\-Error & \intnum{5} & \num{0} & \num{0} & \num{0} & \num{0} & \num{0} & \num{0} \\
  Zero\-Division\-Error & \intnum{5} & \num{100} & \num{20} & \num{80} & \num{100} & \num{20} & \num{80} \\
  NameError & \intnum{3} & \num{0} & \num{0} & \num{0} & \num{0} & \num{0} & \num{0} \\
\end{talltblr}
\end{table}

For each exception,~\autoref{tab:exceptionscounts} shows, for each model, when that particular exception is predicted precisely
(``Strict match''), and when the model predicts that \emph{some} exception
is predicted (``Any Match''). A strict match may occur if one of the heuristic substrings for that \emph{particular} exception is contained in the model output \eg `\texttt{infinite}' for the ground-truth exception  \texttt{TimeoutError}. Unlike infinite recursions (RecursionError), infinite loops are not a runtime exception in Python; thus, predicting the non-standard TimeoutError verbatim is challenging. We found that all correct predictions here are due to `\texttt{infinite}' appearing in the model output, per the heuristic. We later find that with prompt variations, Gemini 3 Pro can indeed predict TimeoutError \emph{verbatim} for up to \pct[0]{19.444} of cases.

\begin{table}[t]
\centering
\caption{Results for frontier models on \cruxevalExc. \textit{Strict Match} requires the generation to contain either the ground-truth exception type, exception message, or a heuristic substring. \textit{Any Match} requires the generation to contain \emph{any} exception type or heuristic substring \eg model output contains "infinite", but ground-truth exception is \texttt{TypeError}.}
\label{tab:exceptions}
\begin{tblr}{
  colspec = {X[l] X[l] X[r] X[r]},
  row{1} = {font=\bfseries},
  hline{1,2,Z} = {0.8pt},
  hline{6,10} = {dashed},
  cell{2}{1} = {r=4}{l},
  cell{6}{1} = {r=4}{l},
  cell{10}{1} = {r=4}{l},
  columns = {font=\small},
  stretch = 0.8,
  rowsep = 0.5pt,
}
  Model & Prompt & Strict Match \% & Any Match \% \\
  GPT-5 Nano & Original & \num{71.9468} & \num{73.156} \\
  GPT-5 Nano & Exceptions & \num{88.4892} & \num{93.6451} \\
  GPT-5 Nano & Type & \num{73.5012} & \num{74.7002} \\
  GPT-5 Nano & Exc. + Type & \SetCell{font=\small\bfseries} \num{89.9281} & \SetCell{font=\small\bfseries} \num{95.9233} \\
  GPT-5.2 & Original & \num{14.7659} & \num{14.7659} \\
  GPT-5.2 & Exceptions & \num{80.5755} & \num{88.6091} \\
  GPT-5.2 & Type & \num{15.5875} & \num{15.7074} \\
  GPT-5.2 & Exc. + Type & \SetCell{font=\small\bfseries} \num{81.295} & \SetCell{font=\small\bfseries} \num{91.1271} \\
  Gemini 3 Pro & Original & \num{48.5612} & \num{50.7194} \\
  Gemini 3 Pro & Exceptions & \num{98.2014} & \num{98.9209} \\
  Gemini 3 Pro & Type & \num{81.8945} & \num{84.4125} \\
  Gemini 3 Pro & Exc. + Type & \SetCell{font=\small\bfseries} \num{98.9209} & \SetCell{font=\small\bfseries} \num{99.6403} \\
\end{tblr}
\end{table}

In light of these two correctness criteria, we find that the performance gap is particularly evident in the 358 occurrences of TypeError exceptions:
Gemini 3 Pro correctly predicts that TypeError exception \emph{per se} occurs just \(\sim\)28\% of the time, while predicting \emph{some} exception \(\sim\)32\% of the time. In most instances of TypeError where GPT-5 Nano predicted \emph{some} exception, it also almost always correctly predicted the specific TypeError, as evidenced by the minimal difference of \pct[1]{2.8} between `any' and `strict' match, \pct[1]{58.82} and \pct[1]{56.02} respectively. Upon further examination, we found that GPT-5 Nano correctly predicted the \emph{exact} exception message for \pct[2]{48.739496} of TypeErrors.

Among the exceptions that occur more
frequently, TypeError appears to be hardest one for all our frontier models\footnote{While
performance for LookupError \& NameError  are worse, we have too few
samples of them to draw any conclusions.}.
In Python, TypeError functions as a catch-all, spanning, by convention, a diverse range of errors, such as argument count mismatches and non-iterable access.
Python experts also use it as a broad, domain-extensible error type.
This diversity likely accounts for both its prevalence in \cruxevalExc and its relative prediction difficulty. 
Whether encountered statically or dynamically,
this difficulty in predicting TypeErrors cannot be ignored: 
TypeErrors \emph{will be} encountered when writing \& debugging
code; thus, it would be desirable to improve the ability to predict occurrences
of this error. We begin by first discussing possible reasons of poor performance in predicting exceptions in general, and then turn more specifically to TypeErrors.

First, we note that the \(\sim\)15\% value in  ``Original'' line for GPT-5.2, in~\autoref{tab:exceptions} was produced using the actual, unmodified
\cruxeval prompt~\cite{guCRUXEvalBenchmarkCode2024}. 
Notably, this original prompt doesn't mention exceptions, as you can see if you inspect the \cruxeval prompt, labeled ``Direct Output Prompt'' in~\autoref{fig:prompts}. This lacuna may have
caused the frontier LLMs to perform badly. Instruction-tuned models have been
reported to show sycophancy~\cite{sharma2023towards}, where they follow instructions
literally \& narrowly, ignoring other relevant context. For further clarity, we modified the \cruxeval prompt, adding \underline{{\bf just}} the \textcolor{blue}{blue text} in the ``Direct Output Prompt Type-Strict + Exceptions'' prompt in the lower part of~\autoref{fig:prompts}, to tell the frontier LLMs to consider exceptions.
This text comprises both an instruction and adds a few-shot example.

The results, as shown on the ``Exceptions'' line
in~\autoref{tab:exceptions}, improve dramatically from \(\sim\)15\% for GPT-5.2 to \(\sim\)81\%,
which is in the range of what we note for non-exception-raising inputs; in fact, it is able
predict the presence of \emph{some} exception \(\sim\)88\% of the time. Performance
also improves for Gemini, from \(\sim\)49\% to \(\sim\)98\%, and for GPT-5 Nano, from
\(\sim\)72\% to \(\sim\)88\%. 

\begin{figure}[t]
\caption{The prompts we used: the ``Direct Output Prompt'' is from the original
\cruxeval paper~\cite{guCRUXEvalBenchmarkCode2024}. The lower prompt adds instructions for \emph{both} \textcolor{red}{type-tracking} and \textcolor{blue}{exception-tracking}. We also experimented with each separately}
\label{fig:prompts}
\begin{promptbox}[prompt:directoutput]{Direct Output Prompt}
You are given a Python function and an assertion containing an input to the function. Complete the assertion with a literal (no unsimplified expressions, no function calls) containing the output when executing the provided code on the given input, even if the function is incorrect or incomplete. Do NOT output any extra information. Provide the full assertion with the correct output in [ANSWER] and [/ANSWER] tags, following the examples.

[PYTHON]
def f(n):
    return n
assert f(17) == ??
[/PYTHON]
[ANSWER]
assert f(17) == 17
[/ANSWER]

[PYTHON]
def f(s):
    return s + "a"
assert f("x9j") == ??
[/PYTHON]
[ANSWER]
assert f("x9j") == "x9ja"
[/ANSWER]

[PYTHON]
\placeholder{\{code\}}
assert f(\placeholder{\{input\}}) == ??
[/PYTHON]
[ANSWER]
\end{promptbox}

\begin{promptbox}[prompt:direct_output_exc_type]{Direct Output Prompt (Type-Strict + Exceptions)}
You are given a Python function and an assertion containing an input to the function. Complete the assertion with a literal (no unsimplified expressions, no function calls) containing the output when executing the provided code on the given input. \textcolor{red}{STRICTLY ANALYZE both the TYPES and VALUES of every variable. Consider how type-specific operations and implicit type conversions affect the final result.} \textcolor{blue}{If the code raises an exception, complete the assertion with the exception type and message.} Do NOT output any extra information. Provide the full assertion with the correct output in [ANSWER] and [/ANSWER] tags, following the examples.

[PYTHON]
def f(n):
    return n
assert f(17) == ??
[/PYTHON]
[ANSWER]
assert f(17) == 17
[/ANSWER]

\textcolor{blue}{[PYTHON]
def f(n, p):
    return n[p]
assert f([1, 2, 3], 4) == ??
[/PYTHON]
[ANSWER]
assert f([1, 2, 3], 4) == "IndexError: list index out of range"
[/ANSWER]}

[PYTHON]
\placeholder{\{code\}}
assert f(\placeholder{\{input\}}) == ??
[/PYTHON]
[ANSWER]
\end{promptbox}
\end{figure}

We now focus on the high prevalence
of Type-related exceptions in~\autoref{tab:exceptionscounts}. We hypothesized that models might be too narrowly focused
on predicting an output \emph{value}, and ignoring \emph{types}; this seems
specially undesirable for a dynamically-typed language like Python, where the interpreter tracks
\emph{both types and values}. We therefore tried additional prompting \textbf{\ul{just}} for
type-tracking, shown in the ``DOP Type-Strict + Exceptions'' prompt in \textcolor{red}{red text}.
We find that type-track prompting does improve performance, although never
as much as exception-track prompting; it helped a lot with Gemini, but less so for the two
GPT models (`Type'' line,~\autoref{tab:exceptions}). Delving into the data,
we find that the type-track prompting doesn't \emph{specifically
 help improve TypeError identification a great deal}; in fact, for the GPT models, 
it helps ValueError identification more! Interestingly, the Exception
 track prompting helps find TypeErrors \emph{much more} than Type-tracking, in  
 all models.

We also tried combining type-track prompting with exception-track
prompting:  this is the ``DOP Type-Strict + Exceptions'' prompt including both the blue and red additions.
As seen in the ``Exc. + Type'' line, their combination always works best, although the improvements over the ``Exceptions''-only prompt are modest.

While the combination prompt improves performance \emph{for all}
the frontier models, it does raise the question as to whether
this improvement is only manifest for the exception-raising cases: would the
combination prompt actually help for normally-running inputs as well? To study this, we used the combination prompt on all our normally-running inputs (both the original \emph{and} perturbed inputs). The results are in~\autoref{tab:promptvariations}.

For both GPT-5 Nano and Gemini 3 Pro, 
when we compare the change from ``Original'' prompt to the combined
``Exc. + Type'' prompt, we see
improved performance for the original (\(C_o\)), the perturbed (\(C_{\tilde{o}}\)) and
both together (\(C_{o \, \cup \, \tilde{o}}\)); we also see improved performance
on the PSR score (the fraction of programs where every prediction is correct).
Not surprisingly, improvements are larger for GPT-5 Nano than for Gemini 3, which already performed almost perfectly with the Original prompt. 
We do notice a substantial performance drop-off in GPT-5.2 when we use the combination prompt, from \(\sim\)81\% to \(\sim\)76\%. 
To investigate this 5\% drop in accuracy, we inspected the results to check
if GPT 5-2 is sycophantically predicting more exceptions in response to the
combination prompt. We found that in about \(\sim\)4\% of the cases where it originally
was correctly predicting the right values, it was \emph{indeed} erroneously predicting
exceptions. The causes for the rest of the errors remain unclear. As noted earlier in~\autoref{sec:rq1-results}, we hypothesize the worse behavior
of GPT-5.2 relative to GPT-5 Nano may be due to the latter being better
regularized.

\begin{table}[t]
\centering
\caption{Evaluating the combined Type \& Exception Tracking prompt on the (original
and perturbed) inputs that run without exceptions}
\label{tab:promptvariations}
\begin{tblr}{
  colspec = {X[l,2] X[l] X[r] X[r] X[r] X[r] X[r]},
  row{1} = {font=\bfseries},
  hline{1,2,Z} = {0.8pt},
  hline{5} = {dashed},
  columns = {font=\small},
  stretch = 0.8,
  rowsep = 0.5pt,
  colsep = 0.8pt
}
  Model & Prompt &
\(C_{o}\) (\textuparrow) & 
\(C_{\tilde{o}}\) (\textuparrow) & 
\(C_{o \, \cup \, \tilde{o}}\) (\textuparrow) & 
PSR (\textuparrow) \\

GPT-5 Nano & Original & \num{96.7836} & \num{96.6082} & \num{96.6241} & \num{82.7485} \\
  GPT-5.2 & Original & \SetCell{font=\small\bfseries} \num{99.1228} & \num{79.6053} & \num{81.3796} & \num{56.1404} \\
  Gemini 3 Pro & Original & \num{99.1045} & \SetCell{font=\small\bfseries} \num{99.3582} & \SetCell{font=\small\bfseries} \num{99.3351} & \SetCell{font=\small\bfseries} \num{97.1642} \\
  GPT-5 Nano & Exc.+Type & \num{97.076} & \num{96.8713} & \num{96.89} & \num{83.6257} \\
  GPT-5.2 & Exc.+Type & \num{76.1696} & \num{75.9795} & \num{75.9968} & \num{50.2924} \\
  Gemini 3 Pro & Exc.+Type & \SetCell{font=\small\bfseries} \num{99.4152} & \SetCell{font=\small\bfseries} \num{99.3275} & \SetCell{font=\small\bfseries} \num{99.3355} & \SetCell{font=\small\bfseries} \num{97.2222} \\
\end{tblr}
\end{table}

\subsection{Threats to Validity}
\paragraph{Internal Threats}{Performance was highly dependent on instruction; for example, GPT-5.2 improved from \pct{15} to \pct{81} when prompted to track exceptions. Instruction-tuned models may follow prompts too literally, often failing to predict exceptions because the original prompt did not mention them. Both examples illustrate how the exact choice of prompt can significantly affect outcomes due to \emph{prompt sensitivity}. Additionally, the choice of \emph{model configuration} and sampling hyperparameters may affect results.
While our usage of type-aware mutations aims to create a dense neighborhood of local inputs for each program, the approach may introduce latent \emph{distributional biases} in perturbed inputs  \eg longer input sequences, that have a causal relationship with model outcomes. 
Categorical accuracy for rare exceptions, such as \textit{NameError} or \textit{LookupError}, is unreliable due to \emph{sample sparsity}. 
Finally, the use of regular expressions, heuristics, and their specific selection \& implementations for extracting model responses may introduce \emph{measurement error} or misinterpret the model's intended output.}
\paragraph{External Threats}{While \cruxeval is currently the standard academic benchmark for code-reasoning output prediction, it consists of small, synthetically generated, standalone Python functions that are a coarse-grained proxy for program execution, potentially masking the nuances of state management in larger systems.
Thus, to reduce \emph{benchmark bias}, our findings could be strengthened by applying our approach to multiple, diverse, real-world datasets in future work.
By focusing our study on Python due to the choice of benchmark, we acknowledge a \emph{language bias}. Challenges such as predicting \textit{TypeErrors} may not apply to statically-typed languages, and model accuracy \& robustness might differ on other languages under the same perturbations.
The use of proprietary, black-box models such as GPT-5.2 or Gemini 3 Pro is tied to specific API versions and may change over time under \emph{model decay}.
Our findings are constrained by our perturbation methodology. Without a comprehensive study of perturbations, especially with respect to MPTs, our results may not necessarily generalize to other transformations.}
\section{Related Work}
\label{sec:related_work}

\paragraph{Robustness and Model Stability on Code}
Early research evaluated code models against semantic-preserving transformations. \textcite{henkelSemanticRobustnessModels2022} and \textcite{dongMixCodeEnhancingCode2023} used metamorphic testing (MPT) and ``litmus'' transformations to assess and augment model stability on code-captioning and classification tasks, respectively. ReCode~\cite{wangReCodeRobustnessEvaluation2023} studied robustness in code \emph{generation} under docstring and code perturbation. \textcite{weiCoCoFuzzingTestingNeural2023} applied coverage-guided fuzzing to study the reliability of an LLM determining the semantic equivalence of programs. While benchmarks like \textsc{EvalPlus}~\cite{liuYourCodeGenerated2023} provide comprehensive evaluation for generation, our work focuses on the robustness of understanding \emph{given} code.

Recent works have studied deeper semantic perturbations: \textcite{hoodaLargeCodeModels2024} used counterfactual mutations, such as flipping branches, to test conceptual understanding, finding significant sensitivity to control-flow changes. Others have examined specific contexts such as poor readability~\cite{huHowEffectivelyCode2024}, syntactic adversarial attacks in translation~\cite{yangAssessingImprovingSyntactic2025}, and prompt instability in vulnerability detection~\cite{hanPromptingInstabilityEmpirical2026}.  

\textcite{lamCodeCrashExposingLLM2025} study code reasoning robustness only under code perturbation \eg insertion of misleading natural language comments.
Our RQ2 is a partial reproduction of their core results; our RQ2 findings broadly align with and reinforce theirs.  
We generalize the application of perturbation to inputs. 
\emph{Input perturbation} gives rise to both additional valid, and invalid, inputs (which may cause exceptions), allowing us to evaluate dependency on control flow decisions; our RQ1 and RQ3 explore these issues in detail.

Broader exploration of these topics, can be found in surveys by \textcite{asgariMetamorphicTestingDeep2025} on metamorphic testing for code models, and \textcite{song2026large} on the robustness and reasoning failures of LLMs.

\paragraph{Benchmarking Code Reasoning and Execution}
A significant body of work now benchmarks the deeper code-reasoning capabilities of LLMs. We refer to \textcite{cekaHowDoesLLM2025} for a recent survey taxonomizing code reasoning techniques. Broad multi-task suites like \textsc{CodeMMLU}~\cite{nguyenCodeMMLUMultiTaskBenchmark2025} and \textsc{SX-Bench}~\cite{yanSTEPWISECODEXBenchEvaluatingComplex2025} assess general software principles, while \textsc{Live\-Code\-Bench-Exec}~\cite{jainLiveCodeBenchHolisticContamination2025}, like \cruxeval, assesses output prediction ability. \textsc{CruxEval-X}~\cite{xuCRUXEVALXBenchmarkMultilingual2025} and \textsc{CodeSense}~\cite{royCodeSenseRealWorldBenchmark2025} further extend these evaluations to multilingual and real-world repository settings. In contrast, \textsc{REval}~\cite{chenReasoningRuntimeBehavior2025} uses code generation benchmark data to assesses prediction of intermediate states during program execution, and model logical reasoning consistency on code-intelligence tasks of increasing difficulty. 
Finally, specialized tools such as \textsc{ExeRScope}~\cite{liuToolIndepthAnalysis2025a} isolate runtime behavior and dynamic properties. 

\textcite{prenner2025throwbenchbenchmarkingllmspredicting} contribute a benchmark
that evaluates models \emph{specifically focused} on their ability to predict exceptions in failing programs, with a prompt tuned to find exceptions. We study robustness broadly on the output \emph{and} exception prediction task on programs in \cruxeval, finding limitations, notably for Type Exceptions, and
explore prompt engineering to address these problems. 

\textcite{patel2025planning} discuss exception prediction by first building a CFG and then instructing an LLM use the CFG to predict execution results. We study the robustness of LLM code understanding in a setting wherein it lacks
the support of a separate CFG analyzer. \textcite{bieber2022staticpredictionruntimeerrors} contribute a benchmark that evaluates LLMs' ability to statically determine which exceptions \emph{might} be thrown by a program 
on \emph{any} input, thus a more difficult task; robustness to
perturbation was considered in their work. 

Beyond execution tracing, research has moved toward functional equivalence and static analysis. While \textsc{Core}~\cite{xieCoReBenchmarkingLLMs2025} evaluates static information flow, \textsc{EquiBench}~\cite{weiEquiBenchBenchmarkingLarge2025} and \textsc{ProbeGen}~\cite{allamanisDisprovingProgramEquivalence2025} test whether models can determine if two programs are semantically identical. Notably, \textsc{ProbeGen} utilizes the model itself to generate inputs that disprove equivalence, enabling semantic clustering. Finally, recent work has addressed model confidence through calibration frameworks~\cite{spiessCalibrationCorrectnessLanguage2024} to evaluate and improve code reasoning confidence~\cite{wangOpenOysterEmpirical2025}.
\section{Conclusion}
\label{sec:Conclusion}
Do language models robustly understand the meaning of programs? We explore
this question using perturbations of the \cruxeval benchmark, which requires
models to correctly predict outputs for given program-input pairs. We perturb
the \emph{inputs} in \cruxeval using type-aware mutations, and the \emph{programs} using meaning-preserving transforms. Some of the perturbed inputs cause programs to throw exceptions, and we add these to our evaluation.

We find that performance on perturbed
inputs and perturbed programs is generally a bit lower, notably  \emph{substantially}
lower for the frontier GPT-5.2 model. We also find that performance on exceptions using
the given prompt in \cruxeval is also substantially worse, and report some prompting interventions that provide evidence of improvement. Finally, we also find that these exception-related interventions help all models improve performance on the original benchmarks, but actually \emph{worsen} the performance of GPT\nobreakdash-5.2.

Our findings support the conclusion that even frontier models lack robustness on the
output prediction task. Future
research could explore the causes of this brittleness.

\clearpage

\printbibliography


\end{document}